\documentclass{sig-alternate}
\usepackage{graphicx}

\usepackage[]{hyperref}
\hypersetup{
	bookmarks=true,         
	unicode=false,          
	pdftoolbar=true,        
	pdfmenubar=true,        
	pdffitwindow=false,     
	pdftitle={Adaptation and Abstract Runtime Models},
	pdfauthor={Thomas Vogel and Holger Giese},
	pdfsubject={},
	pdfcreator={},
	pdfproducer={},
	pdfkeywords={adaptive software, structural adaptation, model-driven engineering, models at runtime, model synchronization},
	pdfnewwindow=true,
}

\global\copyrightetc{%Copyright \the\copyrtyr\ ACM \the\acmcopyr\ ...\$10.00
\copyright~Authors/ACM, 2010. This is the author's version of the work. It is posted here for your personal use. Not for redistribution. The definitive Version of Record was published in the Proceedings of SEAMS'10.\\ {\tt DOI:}~\url{https://doi.org/10.1145/1808984.1808989}}

\begin{document}

% --- Author Metadata here ---
\conferenceinfo{SEAMS}{'10, May 2-8, 2010, Cape Town, South Africa}
\CopyrightYear{2010}
\crdata{978-1-60558-971-8/10/05}
% --- End of Author Metadata ---

\title{Adaptation and Abstract Runtime Models}

\numberofauthors{1}
\author{
\alignauthor
Thomas Vogel and Holger Giese\\
       \affaddr{Hasso Plattner Institute at the University of Potsdam}\\
       \affaddr{Prof.-Dr.-Helmert-Str. 2-3, 14482 Potsdam, Germany}\\
       \email{\{thomas.vogel$\lvert$holger.giese\}@hpi.uni-potsdam.de}
}

\maketitle

\begin{abstract}
Runtime adaptability is often a crucial requirement for today's complex software systems. Several approaches use an architectural model as a runtime representation of a managed system for monitoring, reasoning and performing adaptation. To ease the causal connection between a system and a model, these models are often closely related to the implementation and at a rather low level of abstraction. This makes them as complex as the implementation and it impedes reusability and extensibility of autonomic managers. Moreover, the models often do not cover different concerns, like security or performance, and therefore they do not support several self-management capabilities at once.

In this paper we propose a model-driven approach that provides multiple architectural runtime models at different levels of abstraction as a basis for adaptation. Each runtime model abstracts from the underlying system and platform leveraging reusability and extensibility of managers that work on these models. Moreover, each model focuses on a specific concern which simplifies the work of autonomic managers. The different models are maintained automatically at runtime using model-driven engineering techniques that also reduce development efforts. Our approach has been implemented for the broadly adopted \emph{Enterprise Java Beans} component standard and its application is presented in a self-healing scenario requiring structural adaptation.
\end{abstract}


\vspace{-1mm}


\vspace{-1mm}


%=============================================================================
\section{Introduction}\label{sec:introduction}\noindent
Runtime adaptability or even self-manageability is often an important requirement for today's complex software systems~(cf.~\cite{Bencomo:2009,GarCHSS04,KramerM2007,McKinley+2004,302181}).
To accomplish self\--ma\-nage\-ment, a \emph{managed system} is monitored and analyzed, and if changes are required, adaptations are planned and executed. These activities are performed by \emph{autonomic managers} that are located externally to a managed system, which supports reusability and extensibility of managers. However, it requires appropriate representations of a running managed system for managers~\cite{Bencomo:2009,GarCHSS04,KephartChess2003}.

For such a representation, several approaches use an architectural model that is closely related to the specific implementation of a system and that covers different, but mostly one addressed concern for self-management, like performance for self-optimization or failures for self-healing.
For example, to ease the connection between a model and a system, the model of Oreizy et al.~\cite{302181} is a one-to-one mapping between implementation classes and model elements, and therefore rather at a low level of abstraction.
Using the approach of Garlan et al.~\cite{GarCHSS04}, one user-defined model is employed to cover all concerns of interest and their case studies primarily target at optimizing the performance and cost of a system.

We argue that it is beneficial to simultaneously have more than one kind of architectural model at runtime, especially for two reasons:
First, a model reflecting a platform-specific or a one-to-one view on a system is rather related to the specific implementation and it does not always provide an appropriate abstraction of the system for autonomic managers. In contrast, additional models abstracting from the underlying platform or to different extents from implementation details are beneficial as they support the reusability of autonomic managers using these models across different managed systems. Moreover, models at higher levels of abstraction can be less complex than lower level ones, which can simplify the work of managers.
Second, each model can focus on a specific concern and therefore on a specific self-ma\-nage\-ment capability. If only one model is employed, it has to cover all required concerns, which increases the model's complexity.
For example, supporting self-optimization, self-healing and self-protection requires considering performance, failure and security concerns.
In contrast, having a model for each addressed concern, the complexity of the specific models can be reduced. Likewise to appropriate abstraction levels, this can simplify the usage of these models by autonomic managers.

Similar to these ideas, but in contrast to their realized approach~\cite{302181}, Oreizy et al. consider different kinds of runtime models, but the considerations remain theoretical~\cite{1370181}. It is unclear whether different kinds of runtime models might be simultaneously employed with the framework of Garlan et al. that however provides many customization options~\cite{GarCHSS04}.

In this paper we propose a model-driven approach that provides multiple architectural runtime models at different levels of abstraction as a basis for monitoring and adapting a running software system. Higher level models abstract from the underlying system and platform leveraging reusability and extensibility of autonomic managers. Moreover, each of them focuses on a specific concern, and together with appropriate abstractions, this simplifies the work of managers.

The different models are maintained automatically at runtime by model-driven engineering techniques, especially the incremental synchronization between models that are based on different metamodels. These techniques considerably ease the development by reducing the development efforts for re\-presenting and maintaining runtime models.

To support monitoring and adaptation, the models are causally connected to the running system, i.e., changes in a model are reflected in the system and vice versa.
Thus, our notion of runtime models corresponds to the one of Blair et al.~\cite{MRTcomputer09}, who consider the causal connection and the representation of a system together with its concerns from problem space perspectives as key properties of runtime models.

The basic idea of using model synchronization techniques for monitoring and adaptation has been presented in a poster session~\cite{Vogel-ICAC09}.
Details about the monitoring part of our approach that covers the monitoring of architectural constraints, performance and failure states through distinct runtime models has been published in~\cite{VogelMRT09}.
This paper presents our solution for the adaptation part and discusses the inherent challenges we had to resolve related to the abstraction involved. This solution facilitates parameter and structural adaptation using model synchronization techniques, and therefore our approach now leverages the whole feedback loop.
Both parts have been implemented for the broadly adopted \emph{Enterprise Java Beans}~(EJB) component standard~\cite{jsr220} and we will illustrate our approach with a self-healing scenario requiring structural adaptation.

The rest of the paper is structured as follows:
Section~\ref{sec:basic-approach} describes our basic approach, the applied model-driven engineering techniques and the monitoring part.
The adaptation part together with our solutions to the adaptation challenges are discussed in-depth in Section~\ref{sec:adaptation}.
An application example is presented in Section~\ref{sec:application} and the implementation of our approach in Section~\ref{sec:implementation}.
After discussing related work in Section~\ref{sec:related-work}, we conclude with Section~\ref{sec:conclusion}.

%==============================================================================
\section{Basic Approach}\label{sec:basic-approach}
\begin{figure}[t]
        \centering
        \includegraphics[keepaspectratio, width=0.8\columnwidth]{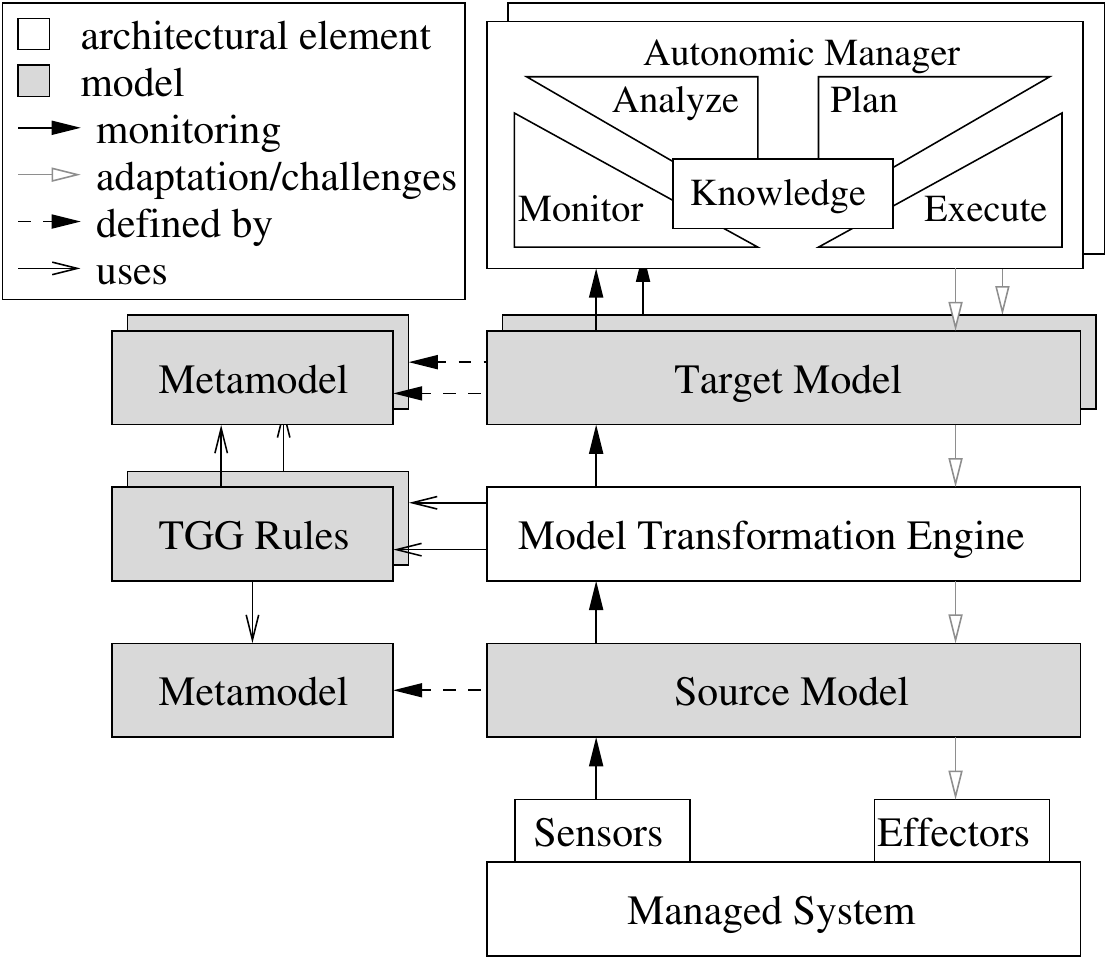}
        \caption{Generic Architecture}
        \label{fig:approach}
       	\vspace{-1mm}
\end{figure}

\noindent
A conceptual view on our basic approach is given by the generic architecture depicted in Figure~\ref{fig:approach}, which extends the control loop concept proposed by Kephart and Chess~\cite{KephartChess2003}.
A \emph{Managed System} provides \emph{Sensors} and \emph{Effectors} that are used to observe and adapt the system, respectively. A runtime representation of the system in the form of a \emph{Source Model} is provided by the sensors and effectors. This model is causally connected to the system. Therefore, adaptation can be conducted by directly using this model.

However, a source model represents all functionalities and concerns of the sensors and effectors. Therefore, it is usually quite complex and related to the solution space and platform of a managed system. Thus, a source model rather provides a view on a system at a low level of abstraction, which could make it laborious to use it as a basis for monitoring and adaptation activities by \emph{Autonomic Managers}.

Therefore, several \emph{Target Models} are derived from a source model at runtime. Each target model abstracts from the source model and it provides a specific view on a managed system required for a certain self-management capability.
As an example, a target model might represent the performance state or failures of a system to address self-optimization or self-healing, respectively. A manager being concerned with self-optimization will use only the target models being relevant for optimizing a system, but does not have to consider target models addressing other capabilities like self-healing. This and appropriate abstractions of models reduce the complexity for specific managers in coping with runtime models and performing their activities.

Thus, target models tend to provide views related to problem spaces of different self-management capabilities and to abstract from the underlying system platform. This supports the reusability of managers that focus on problem spaces shared by different managed systems. Furthermore, as target models can be platform-independent, the kinds of target models used in our approach are rather defined by autonomic managers than by the underlying infrastructure.

Therefore, autonomic managers preferably use target models than a complex source model to monitor or adapt a system. This requires that a target model is causally connected to the source model. Thus, changes in the source model are reflected in target models for monitoring, and vise versa for adaptation.
To maintain different target models at runtime and to realize causal connections between the models, we apply model-driven engineering techniques, especially our \emph{Model Transformation Engine} that incrementally synchronizes models with each other~\cite{GieseHildebrandt08,GW2008}. Details about these techniques, \emph{Metamodels} and \emph{Triple Graph Grammar~(TGG) Rules} are provided in the following section.

As target models abstract from the source model, synchronizing source model changes to a target model for monitoring is not critical. However for adaptation, the opposite direction of propagating target model changes to the source model is critical, as these changes have to be refined in order to be reflected properly in the source model. Our solution to this refinement problem and to other challenges regarding restrictions and appropriate orders of adaptations is the contribution of this paper and discussed in Section~\ref{sec:adaptation}.

%==============================================================================
\subsection{Role of Model-Driven Engineering}\label{sec:basic-approach:foundations}
\begin{figure}[htb]
        \centering
        \includegraphics[keepaspectratio, width=1\columnwidth]{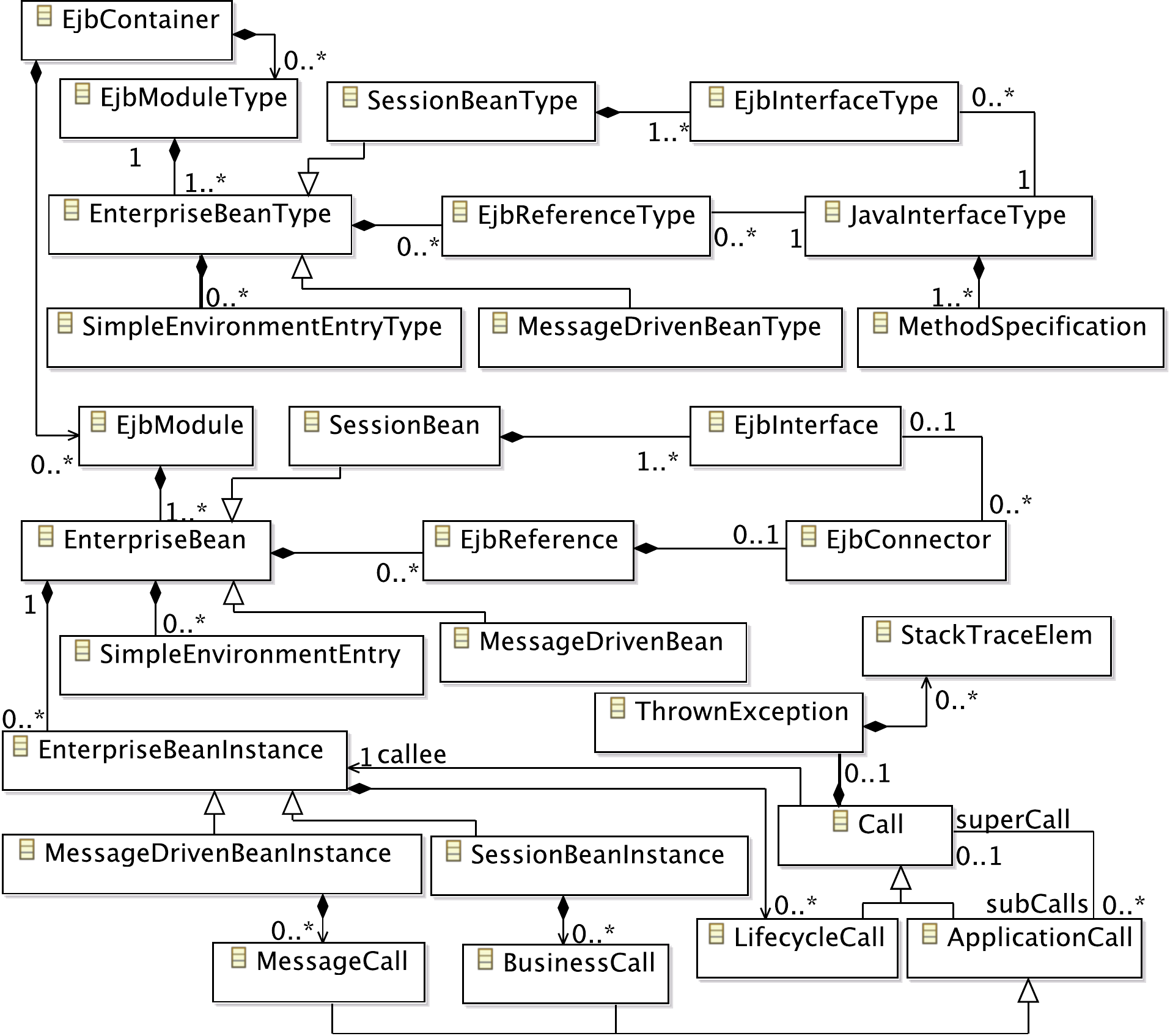}
        \caption{Simplified Metamodel for EJB-based Software Systems}
        \label{fig:source-metamodel}
\end{figure}

\noindent
\emph{Model-Driven Engineering}~(MDE) provides techniques to software system development using models. Each model is defined by a \emph{metamodel} that defines the abstract syntax of a modeling language.
As an example, Figure~\ref{fig:source-metamodel} shows the simplified\footnote{The metamodel depicted in Figure~\ref{fig:source-metamodel} is simplified as it does not show any attributes, operations and enumerations, and it hides some associations. Moreover, elements for concerns like security, transaction, timers or quiescence  are hidden.} metamodel for modeling software systems that are based on \emph{Enterprise Java Beans 3.0}~(EJB)~\cite{jsr220} technology.
EJB is a specification for realizing component-based systems on top of the \emph{Java} programming language. Components are based on \emph{Enterprise Beans} that are either \emph{Message Driven Beans} or \emph{Session Beans}. The former one is accessed through asynchronous message passing, and the latter one provides \emph{interfaces} to access its encapsulated functionality.
Required functionality can be declared for beans through \emph{references} that can be connected to interfaces provided by session beans. Beans can be customized through \emph{Simple Environment Entries} that are kind of configuration properties. As unit of deployment, the EJB standard defines the \emph{EJB module} that must contain at least one bean. The runtime environment for EJB components is called a \emph{container}.

Based on the simplified metamodel depicted in Figure~\ref{fig:source-metamodel}, EJB-based systems can be described at three different layers.
The top layer considers components types that correspond to artifacts from the development phase. These types define the configuration space for a system.
Concrete configurations of a system are instances of these types that can be deployed in a container and are covered by the middle layer.
Finally, the lower layer addresses bean instances and interactions by means of calls among them.

As metamodels are themselves models, a \emph{meta-metamodel} is employed to define them. The \emph{Object Management Group} (OMG) defines the \emph{Meta Object Facility}~(MOF)~\cite{OMGMOF} as the standard meta-metamodel for MDE. The \emph{Eclipse Modeling Framework}~(EMF)\footnote{www.eclipse.org/emf/ (Jan 07, 2010)} provides a partial implementation of the MOF meta-metamodel that is the basis for generating code from metamodels or for model transformation techniques.

\begin{figure}[htb]
        \centering
        \includegraphics[keepaspectratio, width=0.7\columnwidth]{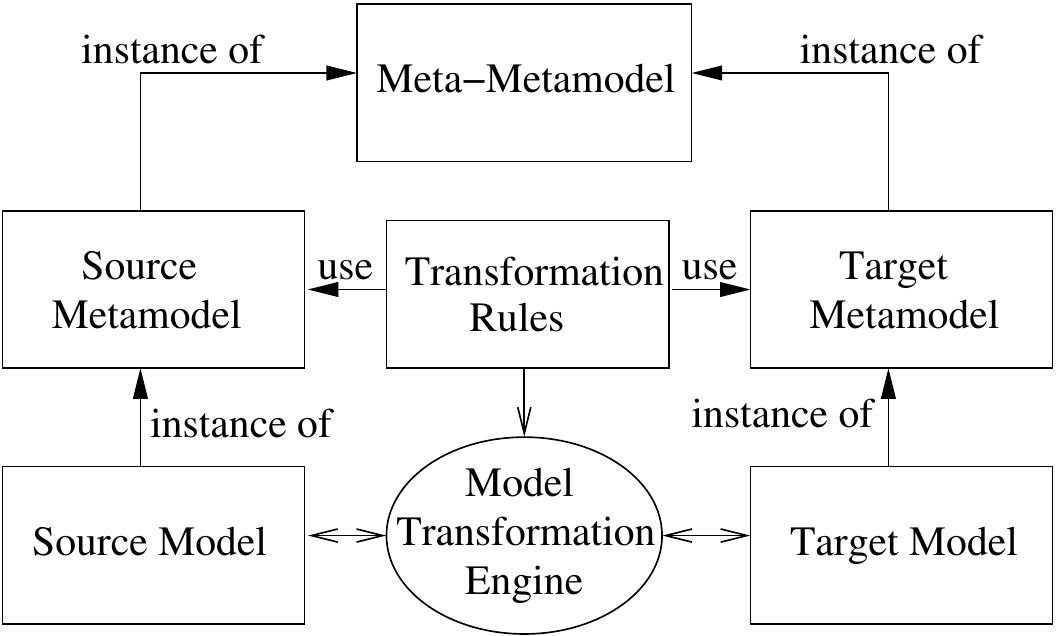}
        \caption{Generic Model Transformation System}
        \label{fig:mts}
\end{figure}

A generic \emph{Model Transformation System} as depicted in Figure~\ref{fig:mts} completely transforms a \emph{Source Model} to a \emph{Target Model} or it synchronizes changes of a \emph{Source Model} to a \emph{Target Model}. If a system has bidirectional transformation capabilities, it also supports the transformation or even the synchronization in the opposite direction, i.e. from a target to a source model. The transformation and synchronization require that both models are instances of potentially different metamodels (\emph{Source Metamodel} and \emph{Target Metamodel}) that share the same \emph{Meta-Metamodel}. Thus, metamodels are user-defined, while the meta-metamodel is usually the MOF. How two models are transformed and synchronized with each other is specified by \emph{Transformation Rules} at the level of metamodels using a textual or graphical language. These rules are used by a \emph{Model Transformation Engine} to perform the transformation or synchronization.

The transformation system used in our approach is based on \emph{Triple Graph Grammars}~(TGGs)~\cite{GieseHildebrandt08,GW2008} and on EMF. It has bidirectional synchronization capabilities that work incrementally. Thus, only the parts of the models are processed by the engine that have been changed and need to be synchronized. To efficiently detect and locate changes in a model, our engine relies on the notification mechanism provided by EMF that reports when a model element has been modified. Consequently, the models have not to be completely scanned if changes have occurred locally. The engine is able to queue notifications and the synchronization can be triggered externally. This enables the synchronization of two models that differ in more than one change.

\begin{figure}[thb]
        \centering
        \includegraphics[keepaspectratio, width=0.95\columnwidth]{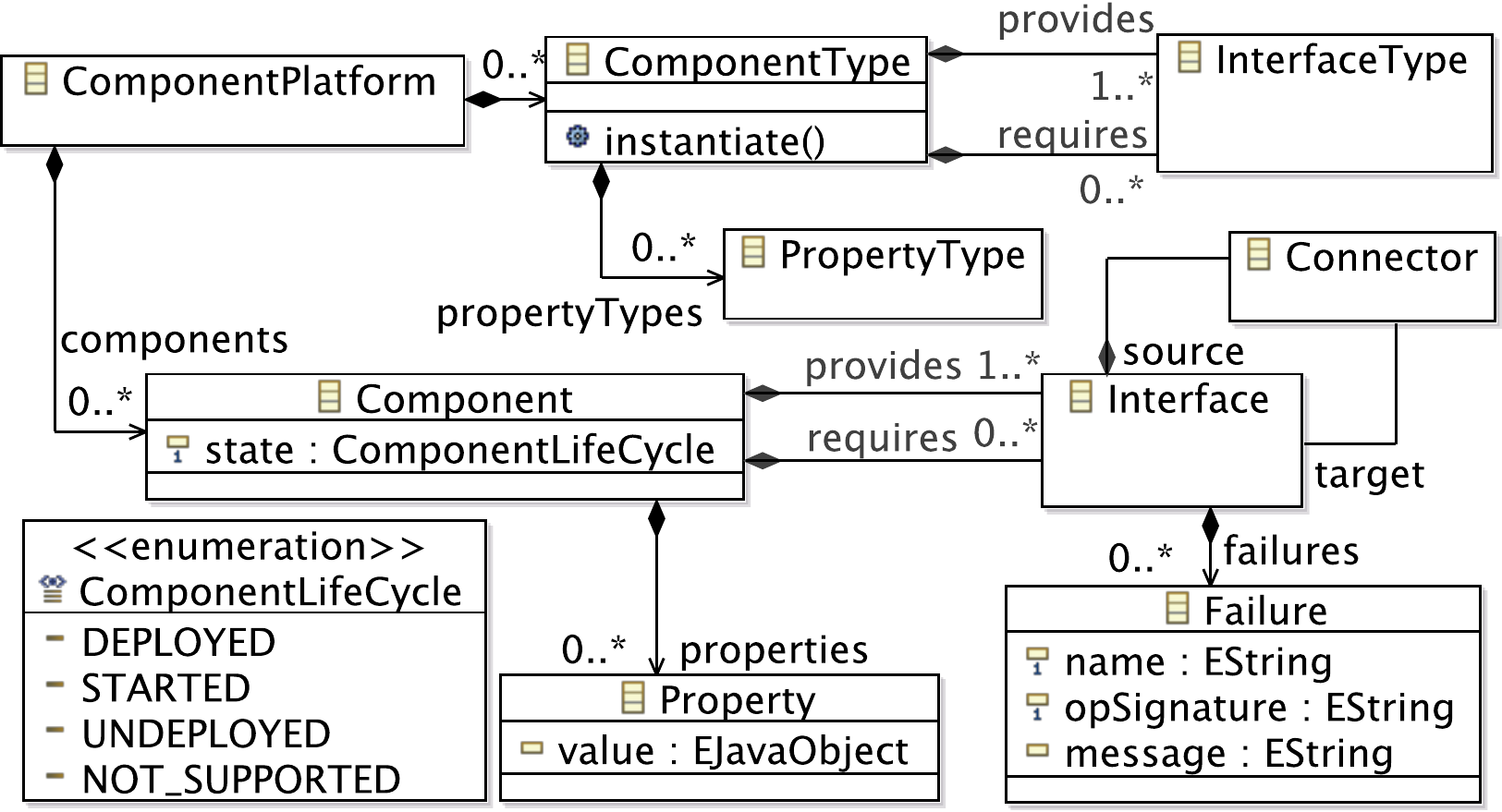}
        \caption{Simplified Metamodel for Component-based Software Systems}
        \label{fig:target-metamodel}
\end{figure}

Before TGGs are presented, we introduce a metamodel for generic component-based software systems, which is depicted in a simplified version\footnote{The depicted metamodel is simplified as several attributes and three associations to navigate from a \emph{Component}, \emph{Interface}, or \emph{Property} to their corresponding types are hidden.} in Figure~\ref{fig:target-metamodel}.
A \emph{ComponentPlatform} is a runtime environment for components. The configuration space for a system is defined by the \emph{ComponentTypes}, their required and provided \emph{InterfaceTypes}, and by \emph{PropertyTypes} that specify options to parameterize components. \emph{Components} conforming to \emph{ComponentTypes} constitute concrete configurations of a system. A component provides at least one \emph{Interface} and it can require \emph{Interfaces} by means of functionality from other components. \emph{Connectors} wire required and provided interfaces, and components are parameterized by setting the \emph{value} attribute of \emph{Properties}. The life cycle of a component can be controlled through its attribute \emph{state}. Finally, the metamodel covers \emph{Failures} that have occurred when using a provided interface.

Using this generic metamodel, EJB-based systems can be described in a platform-independent manner. Therefore, it serves as the target metamodel and the EJB metamodel depicted in Figure~\ref{fig:source-metamodel} as the source metamodel. Hence, instances of these metamodels describing the same system should be synchronized with each other as specified by TGG rules. This requires that for each concept that is represented in both models a TGG rule has to be created. A TGG rule specifies declaratively how a concept in one model is reflected in the other model and vice versa.

Therefore, a TGG combines three conventional graph grammars: one grammar describes a source model, a second one a target model, and a third one a correspondence model. A correspondence model explicitly stores the relationships between corresponding source and target model elements.
The engine uses the correspondence model to efficiently navigate between the source and target models when it checks and reestablishes consistency during the synchronization.
The bidirectional transformation or synchronization is actually performed by operational rules that are generated automatically from declarative TGG rules. There is no need to write any imperative code in addition to the rules, which eases the development of enabling the maintenance of models.

\begin{figure}[htb]
        \centering
        \includegraphics[keepaspectratio, width=0.89\columnwidth]{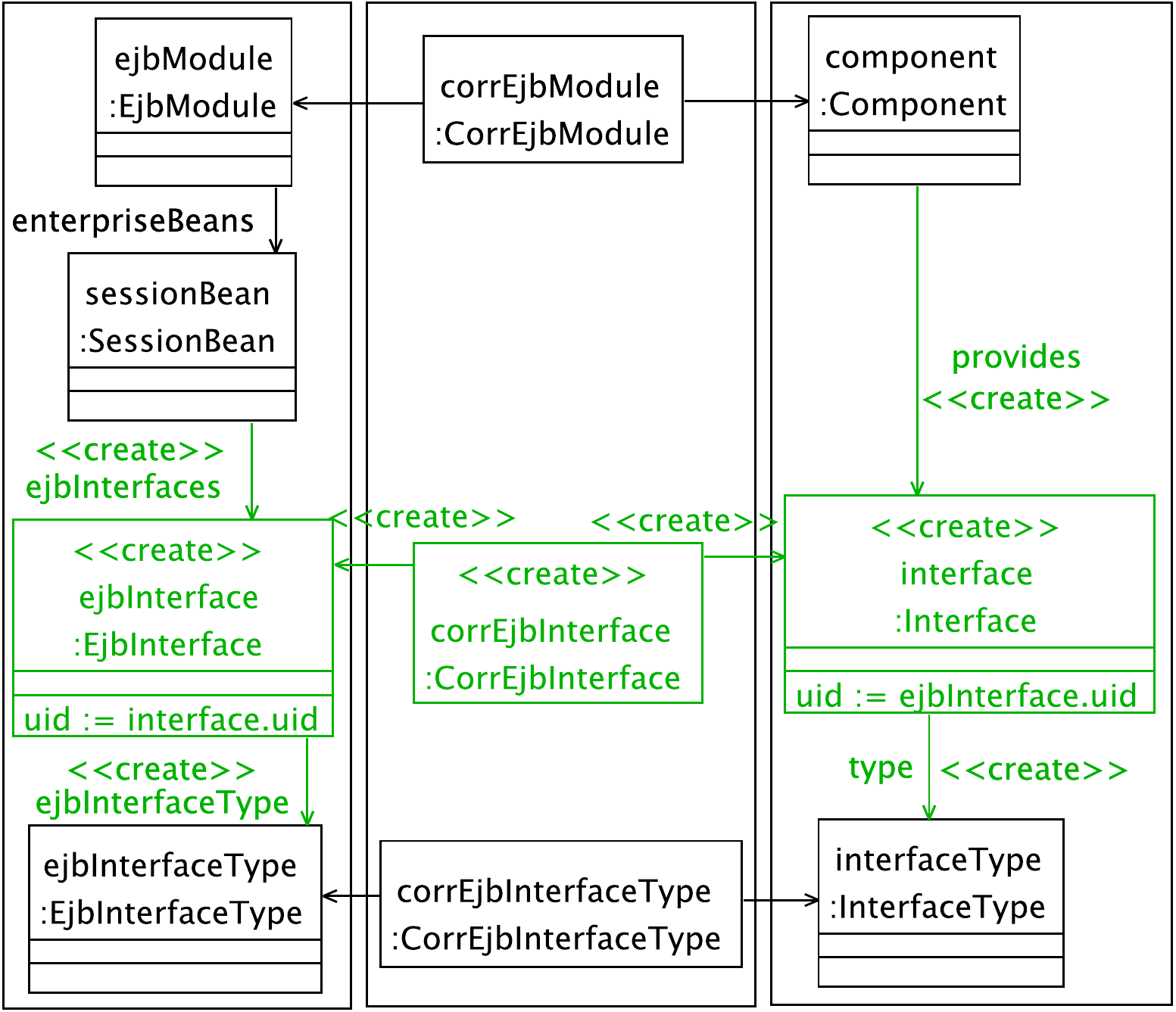}
        \vspace{-1.25mm}
        \caption{Example TGG Rule}
        \label{fig:tggrule}
\end{figure}

Figure~\ref{fig:tggrule} shows a TGG rule that transforms and synchronizes an \emph{EjbInterface} element to an \emph{Interface} element, or vice versa. Model elements on the left refer to the source model, elements in the middle to the correspondence model, and elements on the right to the target model.
The elements having no stereotype form the application context of the rule, i.e., they must already exist in the models before the rule can be applied. The elements being stereotyped with \emph{$\ll$create$\gg$} are created by the rule.
Thus, for each \emph{EjbInterface} provided by a \emph{SessionBean} that is part of an \emph{EjbMo\-dule} in the source model an \emph{Interface} is created in the target model and associated as a provided interface to the \emph{Component} that corresponds to the \emph{EjbModule}. Moreover, a \emph{Corr\-EjbInterface} element as part of the correspondence model is created that stores the mapping between the \emph{EjbInterface} and the \emph{Interface}. Finally, the \emph{Interface} is associated to the \emph{InterfaceType} that corresponds to the \emph{EjbInterfaceType} to which the \emph{EjbInterface} is linked to. Likewise, if an \emph{Interface} is created in the target model, it is transformed or synchronized to an \emph{EjbInterface} in the source model.
This rule also shows how attribute values are synchronized. The \emph{uid} of an \emph{Interface} is directly derived from the \emph{uid} of the \emph{EjbInterface}, and vice versa. Moreover, helper methods or \emph{Object Constraint Language} expressions operating on the source or target models can be used to synchronize attribute values.

This rule exemplifies that not all concepts in one model need to be represented in the other model. A \emph{SessionBean} in a source model is not reflected in the target model and therefore no correspondence model element exists that is connected to a \emph{SessionBean}. Thus, target models can abstract from a source model.

Overall, eleven TGG rules were required to specify the transformation and synchronization between instances of the specific source and target metamodels. As these rules are similar to the rule depicted in Figure~\ref{fig:tggrule}, they are not shown as figures, but the basic bidirectional mapping is as follows.

An \emph{EjbContainer} corresponds to a \emph{ComponentPlatform}, and an \emph{EjbModuleType} to a \emph{ComponentType}. \emph{EnterpriseBeanTypes} and therefore \emph{SessionBeanTypes} and \emph{Message\-DrivenBeanTypes} are not reflected in target models. Thus, the characteristics of a bean type, namely \emph{SimpleEnvironmentEntryTypes}, \emph{EjbReferenceTypes} and \emph{EjbInterfaceTypes} are mapped to \emph{PropertyTypes}, required and provided \emph{InterfaceTypes} respectively, that are associated to the \emph{ComponentType} representing the \emph{EjbModuleType} the bean type is part of.
Likewise, an \emph{EjbModule} is mapped to a \emph{Component}, while the \emph{EnterpriseBeans} being part of an \emph{EjbModule} are not represented in target models. However, the \emph{SimpleEnvironmentEntries}, \emph{EjbReferences} or \emph{EjbInterfaces} of a bean are reflected by \emph{Properties}, required or provided \emph{Interfaces} that are associated to the \emph{Component} representing the \emph{EjbModule} the bean is part of. Connections among required and provided interfaces are reflected in source models as \emph{EjbConnectors} and in target models as \emph{Connectors}.

All source metamodel concepts of bean instances and calls are not considered in the target metamodel, except of \emph{Thrown\-Exceptions}. They are aggregated over all calls invoked on bean instances through a specific interface, and they are attached to the corresponding \emph{Interface} in the target model.

Thus, a target model describes EJB-based systems in a platform-independent manner and at a higher level of abstraction with respect to the source model. A target model provides a black box view on EJB modules and their types, since modules and not single beans are the unit of deployment. Moreover, it maps failures to the architecture in a more amenable manner than the source model. Therefore, instead of the complex and low level source model, such a target model at a higher level of abstraction is used by autonomic managers for monitoring and adapting a system.

%=============================================================================
\subsection{Monitoring}\label{sec:basic-approach:monitoring}\noindent
Using our approach for monitoring requires that changes in a source model, due to changes in a managed system, are reflected in target models. Synchronizing changes in this direction is not critical as target models are at a higher level of abstraction than source models. During synchronization, concepts that are represented in a source model but not in a target model are simply discarded, which causes the intended abstraction. Therefore, changes can be propagated from source to target models without any difficulty.

For example, when synchronizing a source to a target model, both conforming to the corresponding metamodels presented in Section~\ref{sec:basic-approach:foundations}, information about the internal structure of \emph{EjbModules}, i.e. the beans inside a module, are discarded as the \emph{Components} in the target model provide a black box view on modules. This is illustrated by the TGG rule depicted in Figure~\ref{fig:tggrule} where a \emph{SessionBean} in the source model has no corresponding element in the target model.

We presented the monitoring of architectural constraints, performance and failures of a system by using different target metamodels for each concern~\cite{VogelMRT09}. The applied source metamodel is almost the same as the one shown in Figure~\ref{fig:source-metamodel}. Moreover, we have evaluated our approach regarding its development costs and runtime performance. For both of them, the evaluation indicate that our solution is efficient.

%=============================================================================
\section{Adaptation}\label{sec:adaptation}\noindent
Software adaptation can be generally conducted in two ways. \emph{Parameter adaptation} modifies variables of a program and \emph{structural adaptation} changes the software architecture by adding, removing or replacing components and connections among components~\cite{McKinley+2004}.
In our approach, autonomic managers should perform both kinds of adaptation by modifying abstract target models rather than a complex and low level source model. Thus, a manager changes a target model and these changes should be reflected in the source model to adapt the managed system. Therefore, the bidirectional synchronization capabilities of our engine can in principal be used.
For monitoring, it was required that changes in the source model are reflected in the target models (see Section~\ref{sec:basic-approach:monitoring}). For adaptation, the opposite direction is required.

\begin{figure}[htb]
        \centering
        \includegraphics[keepaspectratio, width=0.95\columnwidth]{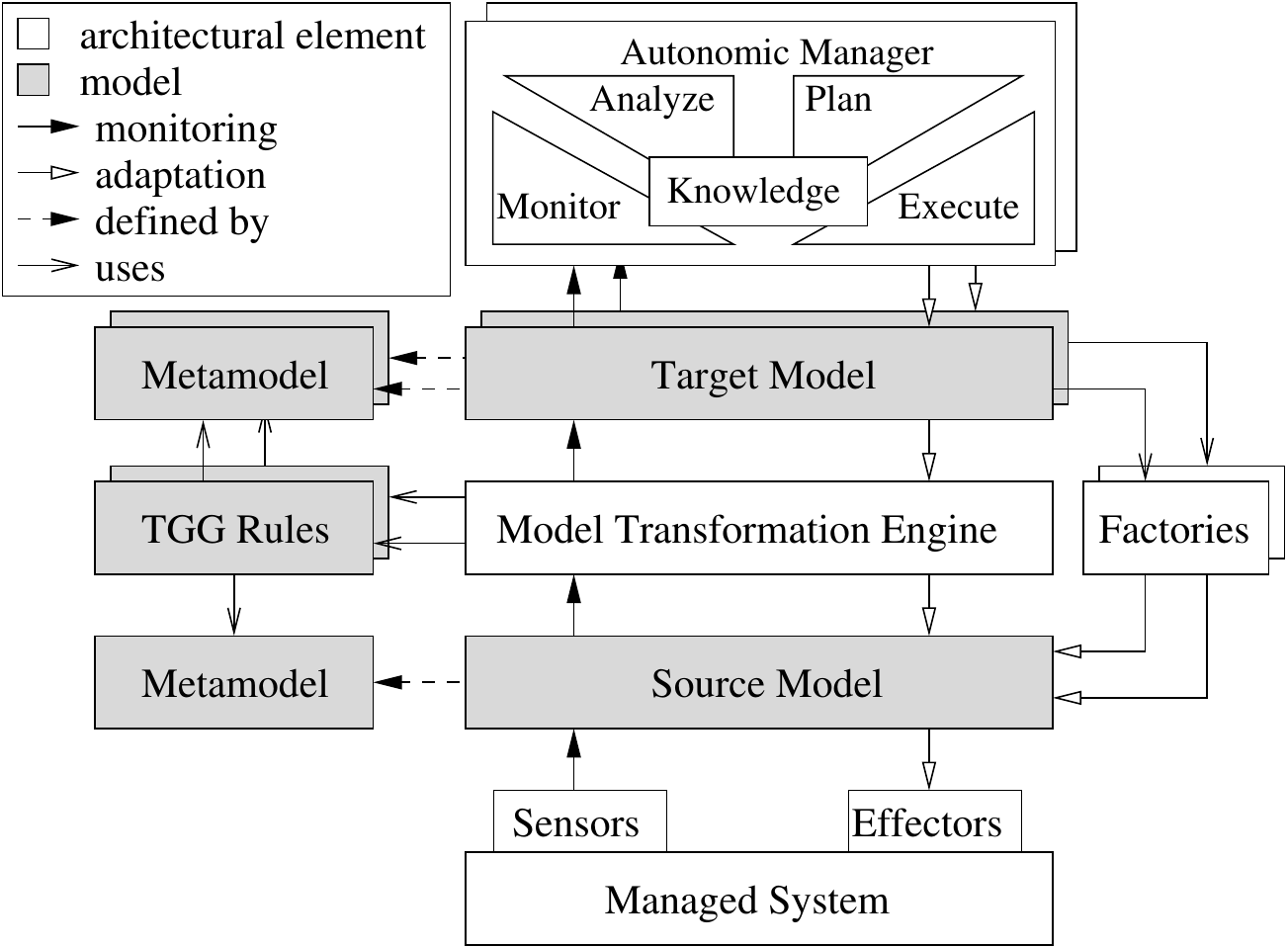}
        \caption{Extended Generic Architecture}
        \label{fig:extended-generic-architecture}
        \vspace{-1mm}
\end{figure}

Until now and with our basic approach described in Section~\ref{sec:basic-approach}, we had no working solution for adaptation. While parameter adaptation is rather feasible without difficulty, we encountered several challenges for realizing structural adaptation. Both kinds of adaptation, the challenges and especially our solution that led to an extended architecture of our approach (Figure~\ref{fig:extended-generic-architecture}) are discussed in the following.

%=============================================================================
\subsection{Parameter Adaptation}\label{sec:basic-approach:param-adaptation}\noindent
Since parameter adaptation modifies program variables, it only requires the synchronization of changed parameter values from a target to a source model. Parameter values are generally of primitive data types, and therefore no complex changes are involved in parameter adaptation. Thus, a target model can only abstract from parameters in the source model by completely omitting them. This implies that adaptable parameters have to be reflected in target models, because managers work on target models and not on the source model (cf. Figure~\ref{fig:extended-generic-architecture}). Consequently, there is usually a bijective mapping between parameters in the source and in the target model, which enables the bidirectional synchronization of parameter values without difficulty.

For EJB-based software systems, parameter adaptation is performed by modifying values for simple environment entries of beans. Using models conforming to the metamodels described in Section~\ref{sec:basic-approach:foundations}, an entry is reflected in the source model as a \emph{SimpleEnvironmentEntry} that is mapped one-to-one to a \emph{Property} in the target model. Thus, the transformation engine can synchronize a changed value of the \emph{Property} attribute \emph{value} to the source model without any problems to finally adapt the parameter in the managed system.

%-------------------------------------------------------------------------
\subsection{Challenges for Structural Adaptation}\label{sec:adaptation:challenges}\noindent
In contrast to parameter adaptation, we encountered at least three research challenges as prerequisites for structural adaptation, which are discussed in the following. Other challenges are sketched in Section~\ref{sec:conclusion} and left for future work.

\subsubsection{Refinement for Adaptation}\noindent
As adaptation is performed on target models that are at higher levels of abstraction than a source model, changes performed by autonomic managers are also at a higher level of abstraction.
If a target model abstracts too much from the source model, the relation between both models is only partial and need not to be bijective. Consequently, a bidirectional synchronization between both models is hindered since information about how to map abstract target model changes to the concrete source model could be missing. Such information has been discarded during monitoring to cause the intended abstraction of target models from the source model (cf. Section~\ref{sec:basic-approach:monitoring}).

If the abstractions are of structural nature, propagating a change from a target to a source model is similar to refining an architecture, which requires additional information that constitute these refinements~\cite{243607,Moriconi+1995}.
Thus, the intended abstractions of target models and the need for refinements of changes can impede structural adaptation based on abstract target models.
In MDE, this problem is generally discussed in round-trip engineering for the bidirectional synchronization of models representing disjoint concepts or concepts at different levels of abstraction~\cite{Hettel_et_al:2008,Stevens2010}.

A refinement problem exists for the synchronization of source and target models conforming to the metamodels described in Section~\ref{sec:basic-approach:foundations}. When a \emph{Component} together with its \emph{Properties}, required and provided \emph{Interfaces} is created in the target model, these changes cannot be synchronized to the source model without additional information. The cause for that is the missing information about \emph{EnterpriseBeans} in the target model as its metamodel does not cover the concept of beans.
The target model only reflects components representing \emph{EjbModules} as black boxes such that the modules' internal structures consisting of beans are hidden. Thus, additional information to create the internal structure of an \emph{EjbModule} in the source model is required during synchronization. This includes information about associating the \emph{SimpleEnvironmentEntries}, \emph{EjbInterfaces} and \emph{EjbReferences}, that correspond to \emph{Properties}, provided and required \emph{Interfaces} respectively, to the appropriate beans being part of the \emph{EjbModule}.
The TGG rule depicted in Figure~\ref{fig:tggrule} exemplifies this case for synchronizing a newly created provided \emph{Interface} in the target model to an \emph{EjbInterface} in the source model. It is undefined, how the \emph{SessionBean} is created in the source model and to which \emph{SessionBean} the newly created \emph{EjbInterface} should be associated to, if the bean contains more than one session bean. Thus, a solution is required that fills the required, but missing information.

We think that this problem cannot be avoided. Otherwise, source and target models would be at the same level of abstraction, which would considerably diminish the benefits of having several target models besides the source model.

\subsubsection{Restrictions to Adaptation}\noindent
Another challenge is the definition of how to interface an autonomic manager with a target model. This means, what kind of changes are allowed on an abstract target model and how a manager performs these changes. As a target model abstracts from a managed system, it can theoretically allow a variety of changes that however could not be executed on the system, for example due to stronger system constraints.

For example, removing a required interface from a component by deleting an interface model element is usually not allowed as the component implementation within the managed system requires the corresponding functionality. Thus, arbitrary changes on a target model could not be supported. Moreover, a target model can be changed by directly adding, removing or modifying model elements, associations among model elements, or attribute values, or by invoking operations provided by model elements. Appropriate means to interface with a model have to be found.
This is similar to the adaptation operators in \emph{Rainbow}~\cite{GarCHSS04}, which determine a set of specific actions a manager can perform on model and system elements for adapting the running system.

\subsubsection{Ordering of Adaptation Steps}\noindent
Finally, the last challenge addresses structural adaptation involving a set of atomic changes or steps that have to be synchronized at once from the target to the source model and even to the system. Thus, rather than propagating each single target model change to the source model and managed system by triggering the model transformation engine, a set of changes should be propagated by one run of the engine.

For example, when using an instance of the target metamodel depicted in Figure~\ref{fig:target-metamodel}, the integration of a new component into a managed system requires at least the following steps. A \emph{Component} element together with its properties, provided and required interfaces has to be created in the target model. The new component must be configured by setting \emph{Property} values and by creating \emph{Connectors} from or to other interfaces of already deployed components. Throughout the adaptation, the life cycle of the new component has to be controlled.
Usually, there exists dependencies among these steps, like a component should only be started when it is configured appropriately. Otherwise, client applications or other components would be able to access the component though it is not completely configured, which is likely to result in failures. Thus, the order of changes matters and the order of changes performed on a target model might even differ from the order of performing corresponding changes to the source model or to the system. As already indicated, not suitable orders might affect the consistency of a system.

%-------------------------------------------------------------------------
\subsection{Our Solution for Structural Adaptation}\label{sec:adaptation-solution}\noindent
This section describes our proposed solution for the challenges presented previously.

\subsubsection{Refinement for Adaptation}\label{sec:adaptation-solution-refinement}\noindent
As a solution for the required refinements to overcome the abstraction gap when transforming or synchronizing structural target model changes to a source model, two approaches have been considered.
First, the missing information for a refinement could be filled with default values. However, this raised questions what meaningful default values for structural aspects are and how to change them subsequently as they would usually not represent the real situation.

Therefore, we applied another approach using \emph{Factories} that are similar to the pattern of \emph{Abstract Factories} using \emph{Factory Methods}~\cite{GammaHJV1995}.
If structural abstractions between source and target model impedes the synchronization of target model changes to the source model, factories are employed to solve the refinement problem.
Rather than conducting changes on the target model by creating or modifying model elements and trying to fill the missing information during synchronization, factories and the monitoring capabilities of our approach are used. Factories can be invoked on target models, while they operate on the source model as depicted in Figure~\ref{fig:extended-generic-architecture}. Thus, autonomic managers only interface with target models and the factory implementations work on the source model where the structural information is sufficiently complete because the abstraction gap does not occur at the source model level. Hence, the intended changes are performed by factories on the source model and afterwards they are synchronized to target models by the transformation engine, which makes them visible for managers.

In our example, a factory is required for creating a \emph{Component} based on a certain \emph{ComponentType} in the target model. The \emph{Component} and its characteristics (\emph{Properties}, required and provided \emph{Interfaces}) are not created manually in the target model and then ineffectively synchronized to the source model, but a factory is used. This factory is invoked through the \emph{instantiate} operation of the specific \emph{ComponentType} in the target model to create or rather instantiate a \emph{Component} of this type. The factory implementation does not introduce any additional information for the refinement as it works on the source model where the information about component types is complete to properly create an instance of a type. Thus, it can create the correct internal structure for an \emph{EjbModule}, i.e., the contained \emph{EnterpriseBeans} with their \emph{EjbInterfaces}, \emph{EjbReferences} etc., in the source model. Finally, these changes are synchronized to the target model by our monitoring capabilities, such that the desired \emph{Component} and its characteristics appear in the target model.

Whether factories directly affect a managed system or not depends on the kind of coupling between the source model and the system. If a source model is closely coupled to a system, each single source model change would be directly executed to the system. In contrast, a set of source model changes could be executed to the system at once, on demand and temporally decoupled from the model changes. Our approach adopts the latter alternative, as it is rather a loose coupling and therefore more flexible.

Summing up, the factories can be seen as a pragmatic extension to the transformation engine as they bypass the engine for the adaptation case when the abstraction between source and target models is too large. However, the engine is used to synchronize changes performed by factories on the source model to target models. This solution gave us first insights towards a fundamental solution extending the capabilities of TGGs that might address the generic challenge of bidirectional model synchronization in MDE.

Nevertheless, the capabilities of the engine to synchronize target model changes to the source model is used for adaptation when components are (re)configured. This is discussed together with our proposed solution to the second challenge.

\subsubsection{Restrictions to Adaptation}\noindent
The second challenge addresses the kind of changes that are allowed on a target model and how they are performed on the model. Our proposed solution is similar to the adaptation operators that have to be defined for specific systems in \emph{Rainbow}~\cite{GarCHSS04}.
Thus, for target models conforming to the metamodel depicted in Figure~\ref{fig:target-metamodel}, we defined the following actions or changes.

As already described, invoking the \emph{instantiate} operation on a \emph{ComponentType} creates a \emph{Component} conforming to this type. The life cycle of a \emph{Component} is controlled by modifying the value of its \emph{state} attribute, which supports deploying, starting, stopping and undeploying a component. Parameter adaptation is supported throughout the life cycle of a component by modifying the value of its \emph{Properties} attribute \emph{value}. A connection among components can be established by creating a \emph{Connector} element and by attaching the \emph{Connector} to a required and provided \emph{Interface}. Deleting a \emph{Connector} element removes the connections among the components. All of these target model changes result in corresponding changes to the source model and therefore to the managed system. Finally, removing a \emph{Component} or a \emph{ComponentType} model element from the target model also removes the corresponding deployable component or respectively the component type from the managed system such that they cannot be used anymore. Other actions or changes performed on a target model are not valid adaptations and therefore they are not permitted.

\subsubsection{Ordering of Adaptation Steps}\noindent
Finally, the last challenge of executing a set of changes on a managed system in an appropriate order that does not affect the consistency of the system is addressed.
This issue is critical as we aim at a loosely and temporally decoupled solution, i.e., each single target model change should not be directly synchronized to the source model and to the system. Therefore, rather a set of target model changes should be propagated to the source model on demand with one synchronization run. Likewise, after each synchronization run, the set of source model changes performed by the engine should be mapped to the system in one run. An advantage of such a decoupled solution is that model changes need not to be executed necessarily on the system, but they could be dismissed if the planned adaptation is not reasonable.

Our transformation engine is able to synchronize in one run two models that differ in more than change. Usually, several target model changes are synchronized in the same order to the source model as they have been conducted on the target model. However, there is no guarantee for that because subsequent changes can overwrite preceding ones that would not be processed by the engine. Thus, the engine might miss intermediate changes that are however relevant from the perspective of a managed system.
A simple example is the life cycle management of a component. If the life cycle state of an undeployed component as part of the target model is changed to deployed and directly afterwards to started, the engine would synchronize these changes to the source model by starting an undeployed component without deploying it in between. This results in a failure as the deployment action has been overwritten by the starting action.
To generally address the issue of ordering a set of changes, three options can be used.

The first option applies to a target model by appropriate modifications of the model and intermediate synchronizations to the source model and managed system. If change $c_1$ should be performed before change $c_2$ on the system, after performing $c_1$ on the target model the synchronization can be triggered. This ensures that a change corresponding to $c_1$ is performed on the source model and on the managed system. Then $c_2$ can be conducted followed by another synchronization. Hence, the order of changes can be controlled by intermediate executions of the synchronization.

Second, the design of the TGG transformation rules can affect the order in which the engine synchronizes several target model changes to the source model in one run. Constraints can be specified in a TGG rule and they serve as a condition that must hold before the rule can be applied. Likewise, a TGG rule has an application context by means of patterns in the source, correspondence and target models (e.g. model elements not being stereotyped with \emph{$\ll$create$\gg$} in Figure~\ref{fig:tggrule}) that must be present for applying the rule. Thus, the interplay of different rules can be utilized. For~example, the changes $c_1$ and $c_2$ are already performed on the target model, but $c_1$ should be executed before $c_2$ on the source model and system. Therefore, the rule for synchronizing $c_2$ has a constraint or application context that is not fulfilled until the rule for $c_1$ has been applied. Thus, the synchronization of $c_2$ is only enabled when $c_1$ has been synchronized before, which results in the desired order of performing $c_1$ before $c_2$. However, this option is at risk of modeling conflicts like cyclic dependencies between different rules.

Finally, the last option takes effect at the causal connection between a source model and a managed system. Therefore, arbitrary changes performed on a source model by the synchronization engine are generally ordered for executing them on a managed system. At first, components that should be stopped are stopped, then connections and components are removed, new components are deployed, connections are created, parameter values are set, and finally, components are started.

Thus, a basic ordering of adaptation steps is supported by the last option, while it is possible to determine more specific orders at design time by modeling appropriate transformation rules (cf. second option) or at runtime by triggering intermediate synchronizations (cf. first option).

%==============================================================================
\section{Application Example}\label{sec:application}
\begin{figure}[htb]
        \centering
        \includegraphics[keepaspectratio, width=0.9\columnwidth]{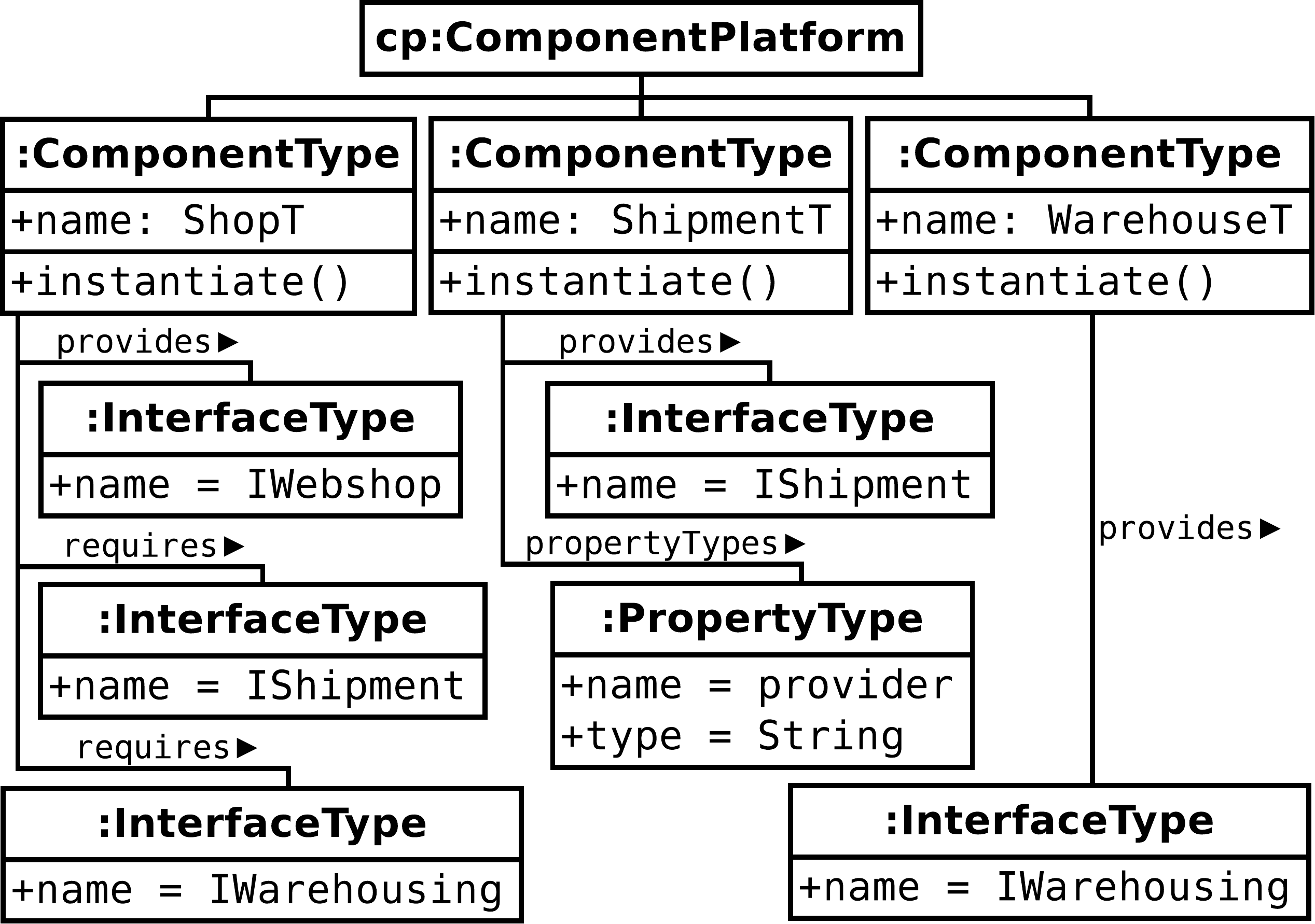}
        \caption{Component Types}
        \label{fig:sample:types}
\end{figure}

\noindent
We demonstrate the application of our approach in a self-healing scenario requiring the replacement of a stateless component\footnote{For simplicity, a stateless component is replaced though basic support for state transfer and quiescence is provided.}. The models depicted in this section are described in the abstract syntax as specified by the target metamodel (cf. Figure~\ref{fig:target-metamodel}).
The application example is a web shop system consisting of three components types that are depicted in Figure~\ref{fig:sample:types}.
The \emph{ComponentPlatform} instance provides the \emph{ShopT}, \emph{ShipmentT}, and \emph{WarehouseT} component types for the core web shop, the shipment of goods, and the ordering of goods from warehouses, respectively. \emph{ShopT} provides the interface type \emph{IWebshop} that specifies the usage of the shop by external client applications and it requires the interface types \emph{IShipment} and \emph{IWarehousing} that are provided by \emph{ShipmentT} and \emph{WarehouseT}, respectively. Finally, \emph{ShipmentT} has a property type that defines the option to configure the shipping \emph{provider} as a \emph{String} value.

\begin{figure}[htb]
        \centering
        \includegraphics[keepaspectratio, width=0.9\columnwidth]{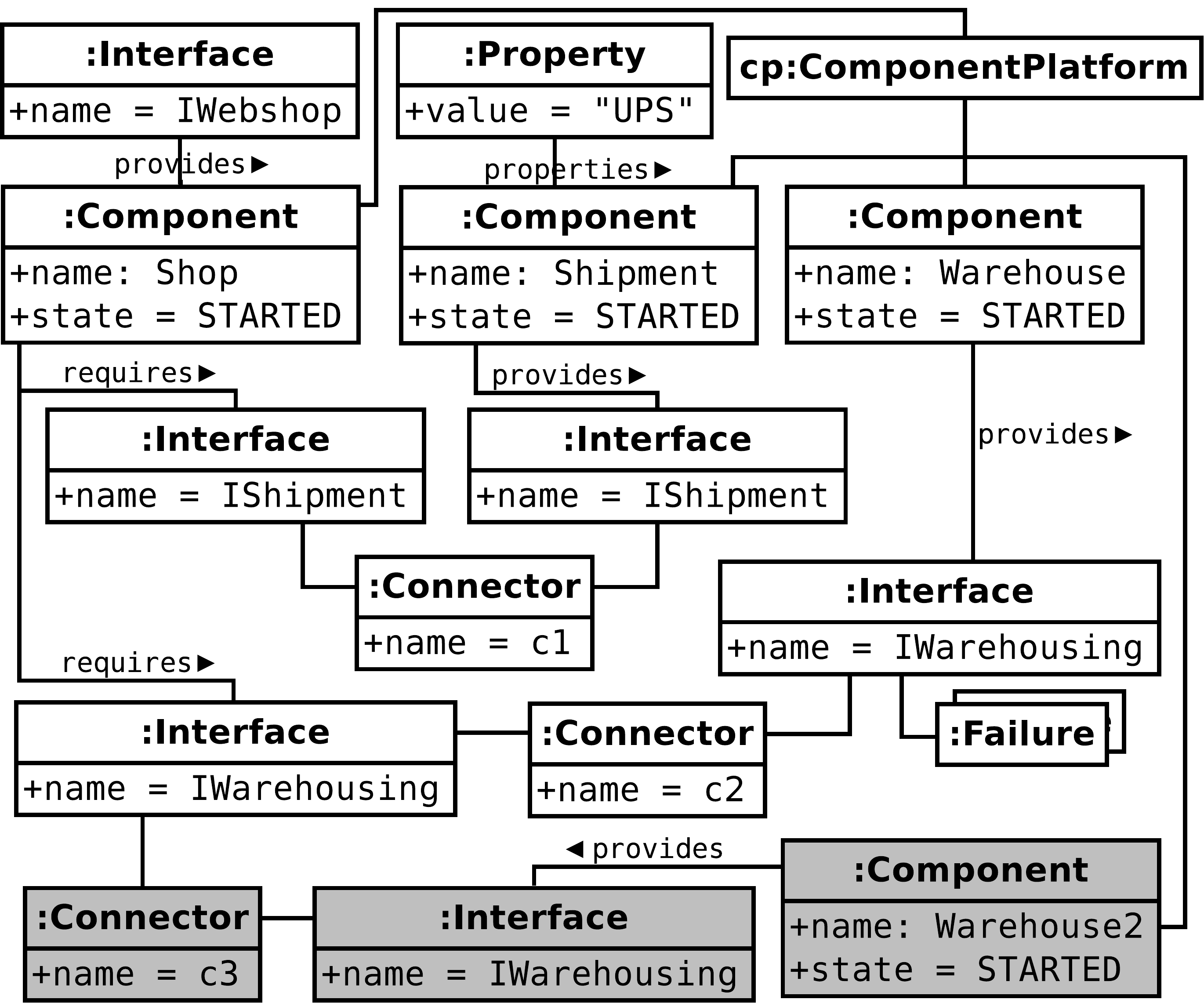}
        \caption{Components and their Configuration}
        \label{fig:sample:configuration}
\end{figure}

A deployed configuration of this shop system is depicted by the white-shaded model elements in Figure~\ref{fig:sample:configuration}. The three component types are instantiated to components, like \emph{ShopT} to \emph{Shop}\footnote{Actually, the elements in Figure~\ref{fig:sample:configuration} and their types in Figure~\ref{fig:sample:types} are part of one model and associated with each other. For illustration, they are depicted separately and the associations can be derived by the values of the name attributes.}, and they are configured. Thus, the \emph{Shop} is wired to the \emph{Shipment} and \emph{Warehouse} components through the connectors \emph{c1} and \emph{c2}, respectively. Finally, the \emph{Shipment} component uses the company \emph{UPS} as a shipping provider, which is configured by the property.

\begin{figure}[htb]
        \centering
        \includegraphics[keepaspectratio, width=0.8\columnwidth]{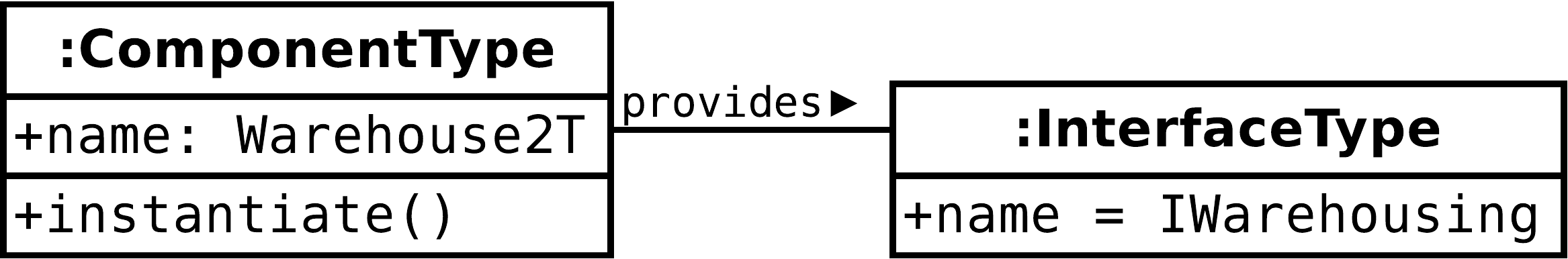}
        \caption{Additional Component Type}
        \label{fig:sample:typesdiff}
\end{figure}

This architectural view on the shop system is a concrete target model\footnote{In contrast to the 21 white-shaded elements of Figure~\ref{fig:sample:types} and \ref{fig:sample:configuration}, the corresponding source model consists of at least 39 nodes. Both numbers do not include the monitored failures.} provided by our approach at runtime. It can be used to locate failures, that have occurred during execution, in the architecture. As defined by the metamodel depicted in Figure~\ref{fig:target-metamodel}, these failures are attached to the specific provided interface whose usages have caused them. This is illustrated in Figure~\ref{fig:sample:configuration} for the \emph{IWarehousing} interface provided by the \emph{Warehouse} component. This information can be used by managers to identify the demand for adaptation to prevent further occurrences of failures.
Therefore, an alternative for the currently used component type \emph{WarehouseT} is required that provides the same interface type and that corrects the faulty behavior.
As soon as new component types are available in the managed system, they appear in the target model due to the monitoring capabilities of our approach. For our example, the \emph{Warehouse2T} component type depicted in Figure~\ref{fig:sample:typesdiff} would enrich the target model depicted in Figure~\ref{fig:sample:types}.

Managers search the target model for component types providing the \emph{IWarehousing} interface type to find with \emph{Warehouse2T} an alternative for the faulty \emph{WarehouseT} component type. In the following, the steps of modifying the target model to replace the \emph{Warehouse} with the \emph{Warehouse2} component are outlined.

Invoking the \emph{instantiate} operation on the \emph{Warehouse2T} model element (Figure~\ref{fig:sample:typesdiff}) creates the \emph{Warehouse2} component and its provided interface \emph{IWarehousing} in the target model (grey-shaded elements in Figure~\ref{fig:sample:configuration}) as described in Section~\ref{sec:adaptation-solution-refinement}. Initially, the new component is in state \emph{UNDEPLOYED} and its provided interface is not attached to any connector.
As the new component has not any required interface or property, it does not need to be configured and it can be directly deployed and started. Therefore, its \emph{state} attribute is set to \emph{DEPLOYED} and then to \emph{STARTED}, each followed by a synchronization to the source model and managed system.
Afterwards, the connector \emph{c2} is removed by deleting the corresponding model element and the new connector \emph{c3} is created an attached to the corresponding required and provided interfaces. These changes are synchronized to the underlying system and the \emph{Shop} is now wired to \emph{Warehouse2} instead of \emph{Warehouse}.
As the \emph{Warehouse} component is not used anymore, it is stopped and undeployed by appropriate modifications of the \emph{state} attribute value, each followed by a synchronization. Finally, the model elements for the \emph{Warehouse} component and \emph{WarehouseT} component type are removed from the target model, which also removes the provided interface and interface type elements, respectively. Synchronizing these changes removes the faulty component and its type from the source model and managed system to prevent any future use of them.

This example showed how a target model is used for identifying adaptation demand and for conducting adaptations, while it abstracts from the underlying platform as it provides a generic component-based view on an EJB-based system.

%==============================================================================
\section{Implementation}\label{sec:implementation}\noindent
Our approach has been implemented on top of the \emph{mKernel} infrastructure~\cite{bruhn2008Comprehensive} that provides sensors and effectors for managing software systems being realized with \emph{Enterprise Java Beans 3.0} (EJB)~\cite{jsr220} technology for the \emph{GlassFish v2}\footnote{https://glassfish.dev.java.net/ (Jan 12, 2010)} application server. EJB-based components are automatically instrumented by \emph{mKernel}, such that management concerns are completely transparent to component developers. The sensors and effectors are provided as an application programming interface~(API) that is not compliant to the \emph{Eclipse Modeling Framework}~(EMF). However, our metamodels and model transformation techniques are based on EMF though they can run decoupled from the \emph{Eclipse} workbench.

Therefore, we developed an EMF-based metamodel for the EJB domain that captures the capabilities of the API and that serves as the source metamodel depicted in Figure~\ref{fig:source-metamodel}. The compatibility to EMF ensures that a source model as an instance of this metamodel can be processed by our model transformation engine.
A source model is causally connected to a running system, which is realized by an event-driven adapter that works incrementally.
Therefore, events dispatched from \emph{mKernel} sensors notifying about changes in the system are processed to update the source model. Likewise, if the source model is changed by the transformation engine, the model emits corresponding events that are mapped to invocations of effectors that actually adapt the system. For both directions, the events can be queued and processed on demand, which results in a loose coupling between the source model and the system. Likewise, the bidirectional synchronization of source and target models can be triggered on demand even if both models differ in more than one change. Thus, our approach achieves a loose coupling between different runtime models and the running system and therefore it can be used in a flexible manner.

To realize structural adaptation, one factory was required. It creates an \emph{EjbModule} in the source model as described in Section~\ref{sec:adaptation}. It also instantiates a corresponding module in the \emph{mKernel} system, but it does not deploy the module and therefore it does not affect the actual system.

%==============================================================================
\section{Related Work}\label{sec:related-work}\noindent
Several approaches address runtime adaptation based on architectural models. In contrast to our approach, they do not provide simultaneously several runtime models at different abstraction levels or for different concerns, like performance or failures to cover different self-management capabilities.

The model of Oreizy et al.~\cite{302181} is as complex as the managed system since it is a one-to-one mapping between model elements and implementation classes. More abstract models are not provided though they theoretically consider hierarchies of models in~\cite{1370181}. Thus, their model can be compared to our complex source model.
Using the framework of Garlan et al.~\cite{GarCHSS04}, it might be possible to simultaneously employ several runtime models at different levels of abstraction or for different concerns, which mainly depends on how the user tailors the framework. However, both case studies presented by Garlan et al. use one model to cover one or more concerns.
Likewise, Caporuscio et al.~\cite{Caporuscio+2007} apply model-based adaptation using one architectural model, but they focus on performance management such that their model does not describe any other concern in addition to performance.
All these approaches~\cite{Caporuscio+2007,GarCHSS04,302181} use some form of architecture description languages and do not apply MDE techniques that can yield to benefits for runtime adaptation~\cite{MRTcomputer09}.

However, MDE approaches also do not provide multiple runtime models, while our approach does. Moreover, they do not work incrementally to maintain a model or they do not apply advanced MDE techniques like model synchronization.
The model of Hein et al.~\cite{MRT07-Hein} is created from scratch out of a system snapshot and it is only used for analysis as adaptation is not supported.
The approach of Dubus and Merle~\cite{MRT06-Dubus} incrementally maintains a model that is however focused on the configuration and deployment of a system and does not cover any additional concern.
Likewise, Morin et al.~\cite{MBJFS09,1555028} update their runtime model incrementally but it only reflects a structural view. However, they use model weaving techniques at runtime to transform models specified by the same metamodel to build new architectures for a managed system.
All approaches discussed before do not use advanced MDE techniques like model synchronization at runtime, except of Morin et al. who apply endogenous transformations, i.e. source and target models share the same metamodel. In contrast, we apply exogenous synchronizations since source and target metamodels are not the same.
Exogenous transformations from software models to analysis models for non-functional concerns are employed by Cortellessa et al.~\cite{conffaseCortellessaMI07} and Sabetta et al.~\cite{SabettaPGM05}. However, they utilize these transformation at design time for analyzing concerns early in the development phase and not at runtime.

Regarding exogenous model transformation or synchronization at runtime, only initial preliminary ideas exist. Song et al.~\cite{Song+2008} use an engine to synchronize models at runtime that however works only offline, i.e., models have to be read from files, which degrades performance. Moreover, their source model does not seem to be maintained at runtime, but rather created on demand from scratch, which might involve non-incremental synchronizations. In contrast, our approach completely works online and incrementally.

% ==============================================================================
\section{Conclusion \& Future Work}\label{sec:conclusion}\noindent
In this paper we presented an approach that uses abstract runtime models and model-driven engineering~(MDE) techniques for adaptation.
In contrast to a complex source model, an abstract target model provides a more appropriate abstraction for autonomic managers and a more specific view for a self-management capability. Both aspects ease the work of managers. Moreover, target models can abstract from a concrete managed system and platform, which supports the reusability and extensibility of managers being able to operate on these models across different systems.
With this work, our basic monitoring approach~\cite{VogelMRT09} has been extended with adaptation capabilities to move towards closed feedback loops.
The approach has been implemented, which was eased for the model synchronization due to MDE techniques, and its application has been shown in an example.

As future work, we will investigate adaptations performed concurrently by different managers on different target models. Since each manager aims at a specific concern, adaptations can interfere with each other. This requires some form of coordination among managers to address interactions between different models, respectively their concerns.
Finally, a distributed setting for our approach is considered.

% ===================================================================

%
\end{document}